\documentclass[fleqn]{article}
\usepackage{espcrc2}
\usepackage{psfig}
\usepackage{graphicx}

\newcommand{\y}{Y(4260)}
\newcommand{\BR}{{\cal B}}
\newcommand{\jpc}{J^{PC}}

\newcommand{\piz}{\pi^0}
\newcommand{\etap}{\eta^{\prime}}

\newcommand{\psp}{\psi^{\prime}}
\newcommand{\pspp}{\psi^{\prime \prime}}
\newcommand{\jpsi}{J/\psi}
\newcommand{\EE}{e^+e^-}
\newcommand{\MM}{\mu^+\mu^-}
\newcommand{\pp}{\pi^+\pi^-}
\newcommand{\kk}{K^+K^-}
\newcommand{\ddb}{D\overline{D}}
\newcommand{\ppjpsi}{\pi^+\pi^- J/\psi}
\newcommand{\beq}{\begin{equation}}
\newcommand{\eeq}{\end{equation}}
\newcommand{\bbcol}{\mbox{BaBar Collaboration}}
\def\eref#1{(\ref{#1})}
\def\Journal#1#2#3#4{{#1} {\bf #2} (#4) #3}

\def\NIMA{Nucl. Instrum. Methods A}
\def\NPB{Nucl. Phys. B}
\def\PLB{Phys. Lett. B}
\def\PRL{Phys. Rev. Lett.}
\def\PRD{Phys. Rev. D}

\title{\boldmath Determining the upper limit of $\Gamma_{ee}$ for the $\y$}
\author{X.H.~Mo \address[IHEP]{Institute of High Energy Physics, 
P.O.Box 918, Beijing 100049, China}\thanks{Supported by National Natural 
Science Foundation of China
(10491302,10491303), 100 Talents Program of CAS (U-25), and
the Knowledge Innovation Project of CAS (U-612(IHEP)).}, 
G.~Li \addressmark[IHEP]$^,$\address[CCAST]{China Center of Advanced
Science and Technology, Beijing 100080, China},
C.Z.~Yuan \addressmark[IHEP], 
K.L.~He \addressmark[IHEP], H.M.~Hu \addressmark[IHEP], 
J.H.~Hu \addressmark[IHEP]$^,$\address[CCAST]{College of 
Physics and Information Technology, Guangxi Normal University, 
Guilin 541004, China}, P.~Wang \addressmark[IHEP], 
Z.Y. Wang \addressmark[IHEP]}

\date{\today}
\begin{document}

\begin{abstract}
By fitting the $R$ values between 3.7 and 5.0~GeV measured by the
BES collaboration, the upper limit of the electron width of the
newly discovered resonance $Y(4260)$ is determined to be 580~eV/$c^2$ 
at 90\% C.L. Together with the BaBar measurement on $\Gamma_{ee}\cdot
\BR(Y(4260)\to \ppjpsi)$, this implies a large decay width of
$Y(4260) \to \ppjpsi$ final states.
\end{abstract}
\maketitle

\section{Introduction}

Recently, in studying the initial state radiation events, $\EE \to
\gamma_{ISR} \ppjpsi$ ($\gamma_{ISR}$: initial state radiation
photon) with 233~fb$^{-1}$ data collected around $\sqrt{s}=10.58$~GeV, 
the BaBar Collaboration observed an accumulation of events
near 4.26~GeV/$c^2$ in the invariant-mass spectrum of
$\ppjpsi$~\cite{babay4260}. The fit to the mass distribution
yields $125 \pm 23$ events with a mass of $4259\pm
8^{+2}_{-6}$~MeV/$c^2$ and a width of $88\pm
23^{+6}_{-4}$~MeV/$c^2$. In addition, the following product is
calculated \beq \begin{array}{l}
\Gamma(Y(4260) \to \EE) \cdot \BR(Y(4260) \to \ppjpsi)) \\ \\
 = 5.5 \pm 1.0^{+0.8}_{-0.7} {\mbox{ eV/$c^2$}}.
\label{gambr}
\end{array} \eeq

Since the resonance is produced in initial state radiation from
$\EE$ collision, its quantum number $\jpc=1^{--}$. However, this
new resonance seems rather different from the known charmonium
states with $\jpc=1^{--}$ in the same mass scale, such as
$\psi(4040)$, $\psi(4160)$, and $\psi(4415)$. Being well above the
$\ddb$ threshold, instead of decaying predominantly into
$D^{(*)} \bar{D}^{(*)}$ final states, the $Y(4260)$ shows strong coupling 
to the $\ppjpsi$ final state. So this new resonance does not seem to be a
usual charmonium state. The strange
properties exhibited by the $Y(4260)$ have triggered many
theoretical discussions~\cite{ebert}-\cite{ywmy}.

One suggestion is that the $Y(4260)$ is the first orbital
excitation of a diquark-antidiquark state
($[cs][\bar{c}\bar{s}]$)~\cite{ebert,maiani}. By virtue of this scheme,
the mass of such a state is estimated to be 4.28~GeV/$c^2$, which
is in good agreement with the observation. A crucial prediction of
the scheme is that the $Y(4260)$ decays predominantly into $D_s
\bar{D}_s$.

Another opinion favors a hybrid explanation~\cite{zhusl,kou,close,luoxq}.
In the light of the lattice inspired flux-tube model, the
calculation shows that the decays of hybrid meson are suppressed
to pairs of ground state $1S$ conventional mesons
\cite{isgur,closeao}. This implies that decays of $Y(4260)$ into
$D\bar{D}$, $D_s \bar{D}_s$, and $D_s^* \bar{D}_s^*$ are
suppressed whereas $D^* \bar{D}$ and $D^*_s \bar{D}_s$ are small,
and $D^{**} \bar{D}$, if above threshold, would dominate ($P$-wave
charmonia are denoted by $D^{**}$). So it is interesting to search
for the possible decay of $Y(4260) \to D_1(2420) \bar{D}$.

The third interpretation we wish to mention is provided by
Ref.~\cite{llanes}, which suggests that the $\y$ is
the second most massive state in the charmonium family. The author
ascribes the lack of $Y(4260)$ in $\EE \to$ hadrons to the
interference of $S$-$D$ waves, and also estimates \beq
\Gamma(Y(4260) \to \EE) \simeq 0.2-0.35 \mbox{ keV/$c^2$ }.
\label{sygee} \eeq

Besides the above interpretations, there are other kinds of
proposals. The lattice study in Ref.~\cite{chiutw} suggests that the
$\y$ behaves like a $D_1 \bar{D}$ molecule. In Ref.~\cite{qiaocf}, 
it is proposed that the new state might be a baryonium, containing charms, 
configured as $\Lambda_c$-$\overline{\Lambda}_c$. In Ref.~\cite{liux}, 
the $Y(4260)$ is considered as a $\rho$-$\chi_{c1}$ molecule while in
Ref.~\cite{ywmy}, the $Y(4260)$ is considered as an
$\omega$-$\chi_{c1}$ molecule. However, all aforementioned
speculations need further experimental judgment.

Most recently, CLEOc collected 13.2~pb$^{-1}$ data at
$\sqrt{s}=4.26$~GeV and investigated 16 decay modes with
charmonium or light hadrons~\cite{cleoy4260}, and the channels
with more than 3$\sigma$ statistical significance are $\ppjpsi$
(11$\sigma$), $\piz \piz \jpsi$ (5.1$\sigma$), and $\kk \jpsi$
(3.7$\sigma$). No compelling evidence is found for any other decay
modes for the $Y(4260)$, nor for the $\psi(4040)$ and
$\psi(4160)$ resonances~\cite{cleoy4260}. These measurements
disfavor the $\rho$-$\chi_{c1}$ molecular model~\cite{liux},
baryonium model~\cite{qiaocf}, and high charmonium state
explanation~\cite{llanes}. So far as other surviving speculations
are concerned, such as charmonium hybrid~\cite{zhusl,kou,close,luoxq},
tetraqark model~\cite{ebert,maiani}, $D_1 \bar{D}$ molecule suggestion,
and $\omega$-$\chi_{c1}$ molecule explanation~\cite{ywmy},
further experimental studies are needed to make more definitive
conclusions.

\begin{figure}[htb]
\includegraphics[width=7.cm]{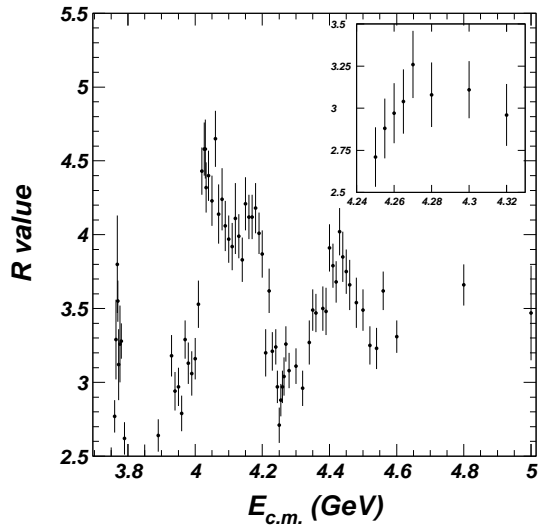}
\caption{\label{fig_rval}$R$ values for $E_{c.m.}$ from 3.7 to
5.0~GeV measured by the BES collaboration~\cite{besr1,besr2}. The
inset shows the $R$ values in the vicinity of the $\y$.}
\end{figure}

Since the $\y$ was observed in $\EE$ annihilation, it is expected
that it contributes to the total hadronic cross section in $\EE$
annihilation (or the $R$ value, in other words). The most recent such
data on $R$ measurements are from the BES
experiment~\cite{besr1,besr2}, as shown in Fig.~\ref{fig_rval}
for $E_{c.m.}(=\sqrt{s})$ from 3.7 to 5.0~GeV. If we look in detail within
the range from 4.25 to 4.30 GeV (refer to the inset of
Fig.~\ref{fig_rval}), it seems there is a bump around 4.27~GeV.
Has this structure a connection with the $Y(4260)$? 
We try to answer this question in this Letter. As there
are other resonances nearby, we shall fit the full spectrum
between 3.7 to 5.0~GeV in order to get the information on the
$\y$.

\section{Fit to the $\psi$-resonances}\label{fitpsisc}

\begin{figure}[htb]
\includegraphics[width=7.cm]{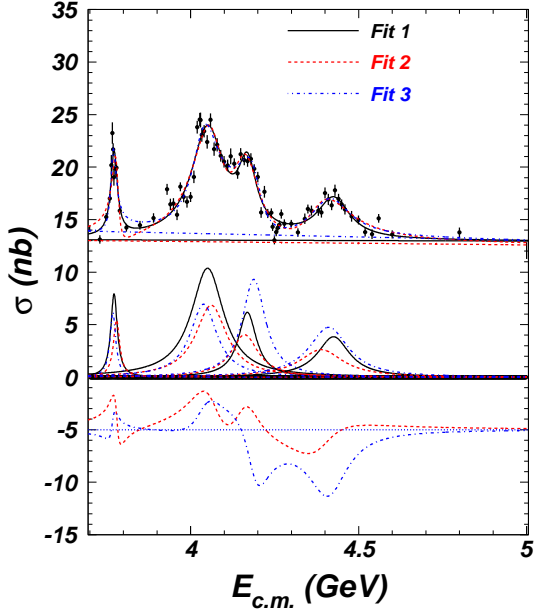}
\caption{\label{fig_psiff}Total hadronic cross section in nb
obtained as $\sigma (\EE \to \mbox{hadrons}) = R \cdot 86.85/s$
($s$ in GeV$^2$) from $R$ values in Refs.~\cite{besr1,besr2}.
Three sets of fit results, Fit 1, Fit 2, and Fit 3 correspond to
three cross section forms $\sigma_1$, $\sigma_2$, and $\sigma_3$
as described in text. Two interference curves have been moved
downward by 5~nb for display purposes; the dashed line at $-5$~nb
corresponds to zero cross section in the fit.}
\end{figure}

The $R$ values~\cite{besr1,besr2} used in this analysis were
measured with the Beijing Spectrometer (BESII), which is a
conventional solenoidal detector expounded in
Ref.~\cite{besdtk}. In the analysis below, the $R$ values are
converted into a cross section $\sigma (\EE)$ by
multiplying the Born order cross section of $\EE\to \mu^+\mu^-$.
These cross sections are plotted in Fig.~\ref{fig_psiff}. There
are clear peaks of $\psi(3770)$, $\psi(4040)$, $\psi(4160)$, and
$\psi(4415)$. The data points have been used to obtained the
parameters of the $\psi$-resonances~\cite{sethr}.

To acquire the resonance parameters, we could simply fit the data
set with cross section formula, each resonance with a Breit-Wigner 
 \beq
 \sigma_j(s) = \frac{12 \pi \Gamma^j_h \Gamma^j_{ee} }
 { [s - (M^j)^2]^2 + (M^j \Gamma_t^j)^2 }~,
 \label{secfm}
 \eeq
where $\Gamma_{ee}$, $\Gamma_h$, and $\Gamma_t$ are the mass
independent electronic, hadronic and total widths, respectively,
for a vector resonance of mass $M$ produced in the head-on
collision of $e^+$ and $e^-$. Notice that $\Gamma_{ee} \ll
\Gamma_t$ in our analysis; the approximation $\Gamma_h \approx
\Gamma_t$ is actually adopted hereinafter. The summation of all
the cross sections within the range we are studying is
 \beq
 \sigma_1 (s) = \sum\limits_{j=1}^{4} \sigma_j(s)~,
 \label{sum_sec}
 \eeq
where indices $1$, $2$, $3$, and $4$ denote four
resonances $\psi(3770)$, $\psi(4040)$, $\psi(4160)$, and
$\psi(4415)$, respectively.

In the $\psi$-family resonance region, if we assume that all
two-body $D^{(*)}\bar{D}^{(*)}$ states are decay products of
resonance, and not produced directly in continuum, we could
therefore treat resonance and continuum incoherently.
Nevertheless, for the four wide resonances, they are close and are
expected to have some same decay final states, there must be
interference between any two of the resonances. Therefore the
amplitudes corresponding to each resonance, with the following
form
 \beq T_j(s) = \frac{\sqrt{ 12 \pi \Gamma_h^j
 \Gamma_{ee}^j } } { [s - (M^j)^2] + i M^j \Gamma_t^j }~,
 \label{ampfm}
 \eeq
have to be added coherently to give the total amplitude that, once
squared, will contain interferences of the type $\Re T_i^* T_j$.
If the resonances are quite broad, the interference effect can
also distort the resonance shape, the width might appear broader
or narrower, and the position of the peak can be displaced as
well. In this case, the total cross section is
 \beq
 \sigma_2 (s) =
 \left| \sum\limits_{j=1}^{4} T_j(s) \right|^2,
 \label{sum_amp}
 \eeq
where $T_j(s)$ is given in Eq.~\eref{ampfm}.

So far as the amplitude is concerned, in principle, there are
presumably relative phases between different amplitudes besides
the phase due to complex Breit-Wigner formula itself. So a more
comprehensive total cross section would be the summation of
amplitudes together with an additional phase, viz.
 \beq
 \sigma_3(s) = \left| \sum\limits_{j=1}^{4} T_j(s) e^{-i \phi_j}
 \right|^2.
 \label{sum_phs}
 \eeq
Since what we actually obtain is the squared modulus of amplitudes,
only three relative phases could be detected in practice.

The standard chi-square estimator is constructed as follows
 \beq
 \chi^2 = \sum\limits_{j=1}^{n} \frac{(\sigma^{exp}(s_j) -
 \sigma^{the}(s_j) )^2}
     { (\Delta \sigma^{exp}(s_j))^2},
 \label{chsqfm}
 \eeq
where $\sigma^{exp}(s_j)$ indicates the experimentally measured
cross section at the $j$-th energy point, while
$\sigma^{the}(s_j)$ is the corresponding theoretical expectation
at this energy point, which is composed of two parts
 \beq
 \sigma^{the}(s_j) = \sigma^{res}(s_j) + \sigma^{con}(s_j)~,
 \label{sigfm}
 \eeq
where $\sigma^{con}$ denotes the contribution from continuum.
Since there is little evidence in the data for any substantial
variation of the continuum background within the studied energy region,
we parameterize the continuum cross section with a linear function
 \beq
 \sigma_{bg}(s) = A + B (\sqrt{s} -  3.700), ( s \mbox{ in GeV}^2 ),
 \label{bgfm}
 \eeq
as has been used in Ref.~\cite{sethr}. Here we consider the
continuum contribution as the background for measurement of
resonance parameters, that is $\sigma_{bg}(s) =\sigma^{con}(s)$.

In Eq.~\eref{sigfm} $\sigma^{res}$ denotes the contribution from
resonances. The fit results are displayed in Fig.~\ref{fig_psiff}
where Fit 1, Fit 2, and Fit 3 correspond to the three cross
section forms $\sigma_1$, $\sigma_2$, and $\sigma_3$, as expressed
in Eqs.~\eref{sum_sec}, \eref{sum_amp}, and \eref{sum_phs},
respectively. Although the synthetic curves for three fits are
almost the same, as we expected, the interference effect deforms
each resonance significantly, according to
Fig.~\ref{fig_psiff}, $\psi(3770)$ and $\psi(4040)$ become
narrower while $\psi(4160)$ and $\psi(4415)$ become wider when interference
effects are included. There exist constructive interferences as well as 
destructive ones, and the interference behaviors for amplitudes with and 
without extra phases are also very distinct. The fit results indicate that
$\Gamma_{ee}$ is very sensitive to the fit strategy, and the
largest difference between various amalgamation strategies could
reach 50\%.

Since $\psi(3770)$, $\psi(4040)$, $\psi(4160)$, and $\psi(4415)$
have similar decay features, that is, all decay dominantly to
$D^{(*)}\bar{D}^{(*)}$ final states, we prefer the synthetic
scheme of amplitude with phase which takes into account all
possible interactions between resonances. 

So far as the $\y$ is concerned, the study of the 
$\bbcol$~\cite{babay4260} implies that this new state may not be a 
common charmonium state, whose decay feature is rather distinctive from 
other $\psi$-family members, and this also obtains support from recent 
CLEOc measurements~\cite{cleoy4260}. So it is favorable to 
perform the incoherent addition for the contribution of
the $\y$ to the total cross sections in Eqs.~\eref{sum_sec}, \eref{sum_amp},
or \eref{sum_phs}. 

In addition, two other effects should
be taken into account. First, there are theoretical arguments in some
references~\cite{maiani,zhusl,kou,close} that the $Y(4260)$ may be due
to a threshold effect just as the $\psi(3770)$, so the width of 
the $Y(4260)$ could be energy dependent as follows:
 \beq
\begin{array}{ll}
 \Gamma_{DD} (s) &= \overline{\Gamma}_{DD} \cdot 
\theta(\sqrt{s}-2m_{D^{*\pm}_s}) \times \\
&{\displaystyle \frac{p^3_{D^{*\pm}_s}}{1+(r p_{\scriptstyle D^{*\pm}_s})^2} \cdot \frac{1+(r\overline{p}_{\scriptstyle D^{*\pm}_s})^2}
 {\overline{p}^3_{D^{*\pm}_s}}~}.
 \label{gtsdpd}
\end{array}
 \eeq
Here the subscript ``$DD$'' explicitly denotes the threshold effect, 
$\overline{\Gamma}_{DD}=\Gamma_{DD}(M^2)$, $r$ is the classical interaction 
radius, $p$ is the $D^{*\pm}_s$ momentum, 
$$ p_{D^{*\pm}_s} = \frac{1}{2}  \sqrt{s - 4 m^2_{D^{*\pm}_s}}~; $$
and $\overline{p}$ is the $D^{*\pm}_s$ momentum at resonance peak,
viz.
$$ \overline{p}_{\scriptstyle D^{*\pm}_s} \equiv
 p_{\scriptstyle D^{*\pm}_s} \Big|_{\sqrt{s}=M} = 
 \frac{1}{2} \sqrt{M^2-4 m^2_{\scriptstyle D^{*\pm}_s} }~. $$
Second, the $\y$ has been observed decaying into the $\ppjpsi$ final 
state and this kind of decay can be expressed by the amplitude with the 
form of Eq.~\eref{ampfm} which is different from that of the decay into 
$D_s^{*+} D_s^{*-}$ with threshold effect.
Furthermore, although the present data display the distinctive feature
of the $\y$ from other charmonium resonances, some connections presumably
exist which could lead to the interference effects of the $\y$ 
with other $\psi$-family members. With all these considerations in 
mind, we split the amplitude of the $\y$ into two parts. One is
the amplitude of the threshold part, which is defined as follows:
\beq T_{DD}(s) = \frac{ \sqrt{ 12 \pi \Gamma_{DD}(s) \Gamma_{ee} } } 
{ (s - M^2) + i M \Gamma_t(s)  }~,
 \label{ampdd}
 \eeq
where the total decay width is defined as the summation of two parts :
$$ \Gamma_t(s)=\Gamma_{DD}(s) + \Gamma_{non}~. $$
Here the first term is the energy dependent partial decay width for the
threshold part while the second term is the energy independent partial 
decay width for the non-threshold part (denoted by $\Gamma_{non}$).
We introduce a factor $f$ to indicate the ratio of $\Gamma_{DD}(s)$
to $\Gamma_t(s)$ at the resonance peak, that is
$$ f=\frac{\Gamma_{DD}(M^2)}{\Gamma_t(M^2)}
 =\frac{\overline{\Gamma}_{DD}}{\overline{\Gamma}_t}~. $$
Notice the energy independence of $\Gamma_{non}$, we have
$\Gamma_{non}=(1-f) \overline{\Gamma}_t$, then  
$$ \Gamma_t(s)=\Gamma_{DD}(s) + (1-f) \overline{\Gamma}_t~. $$
In this case, the amplitude of the non-threshold part is given by
the following expression
\beq T(s)=\frac{\sqrt{12\pi (1-f)\overline{\Gamma}_t \Gamma_{ee} } } 
{ (s - M^2) + i M \Gamma_t(s)  }~.
 \label{ampdd}
 \eeq
Without the knowledge of $f$, we first set $f=0.5$ in the study in 
Sect.~\ref{scanfit} and leave the variation effect of $f$ to 
Sect.~\ref{sect_four}.

By virtue of the above discussion, the synthetic cross section within the 
region studied takes the following form
\beq
\sigma(s) = \left| \sum\limits_{j=1}^{4} T_j(s) e^{-i \phi_j}
+ T(s) e^{-i \phi} \right|^2 + \left| T_{DD} (s)\right|^2~.
\label{sysect}
\eeq
Under such a scheme, the upper limit of the production 
of the $\y$ in $\EE$ annihilation is determined, as expounded in the 
following section.

\section{Determination of $\Gamma_{ee}$ of the $Y(4260)$}\label{scanfit}

Various fits to the data tell us that with limited knowledge on
the nature of the resonances and comparatively meager data, we
could only determine the $\Gamma_{ee}$ of the $Y(4260)$ by a scan
method. Specifically, we fix the mass of the $\y$ at
4.259~GeV/$c^2$ measured by the $\bbcol$~\cite{babay4260}, and scan
over the $\Gamma_t \in (20,180)$~MeV/$c^2$ (8~MeV/$c^2$ step) and
$\Gamma_{ee} \in (0,1000)$~eV/$c^2$ (10~eV/$c^2$ step) parameter
space. In order to avoid some grotesque fit results due to random
effect of the $\y$ on the resonances nearby, in the scan
procedure, the lower and upper bounds of the masses (total widths)
are fixed to be 4100 and 4220~MeV/$c^2$ (30 and 250~MeV/$c^2$) and
4350 and 4500~MeV/$c^2$ (30 and 300~MeV/$c^2$) for $\psi(4160)$ and
$\psi(4415)$, respectively.

\begin{figure}[htb]
\begin{minipage}{7cm}
\includegraphics[width=7.0cm,angle=0]{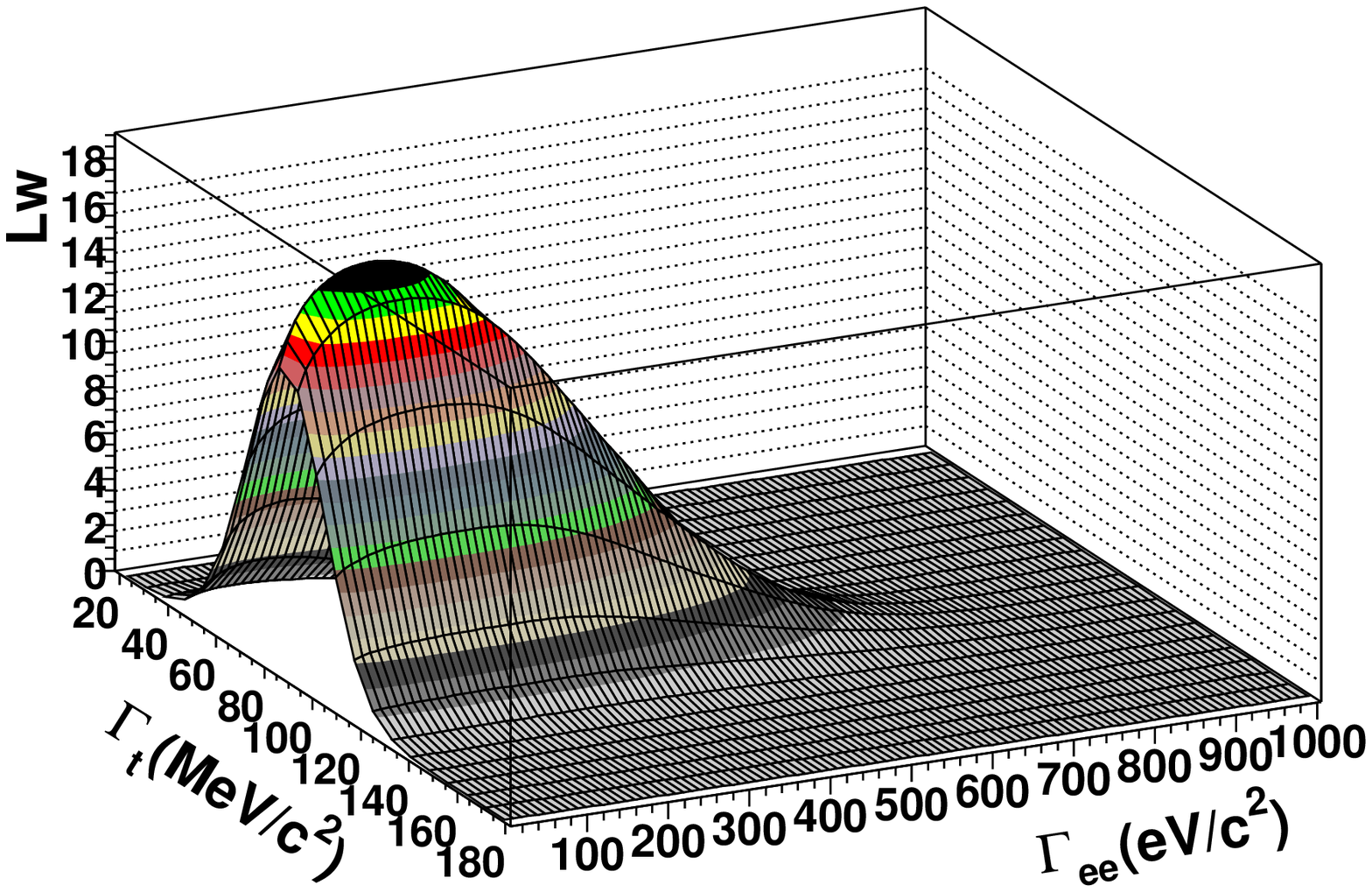}
\centerline{(a) ${\cal L}_w$ w.r.t. $\Gamma_{ee}$ v.s. $\Gamma_t$ }
\end{minipage}
\begin{minipage}{7cm}
\includegraphics[width=7.0cm,angle=0]{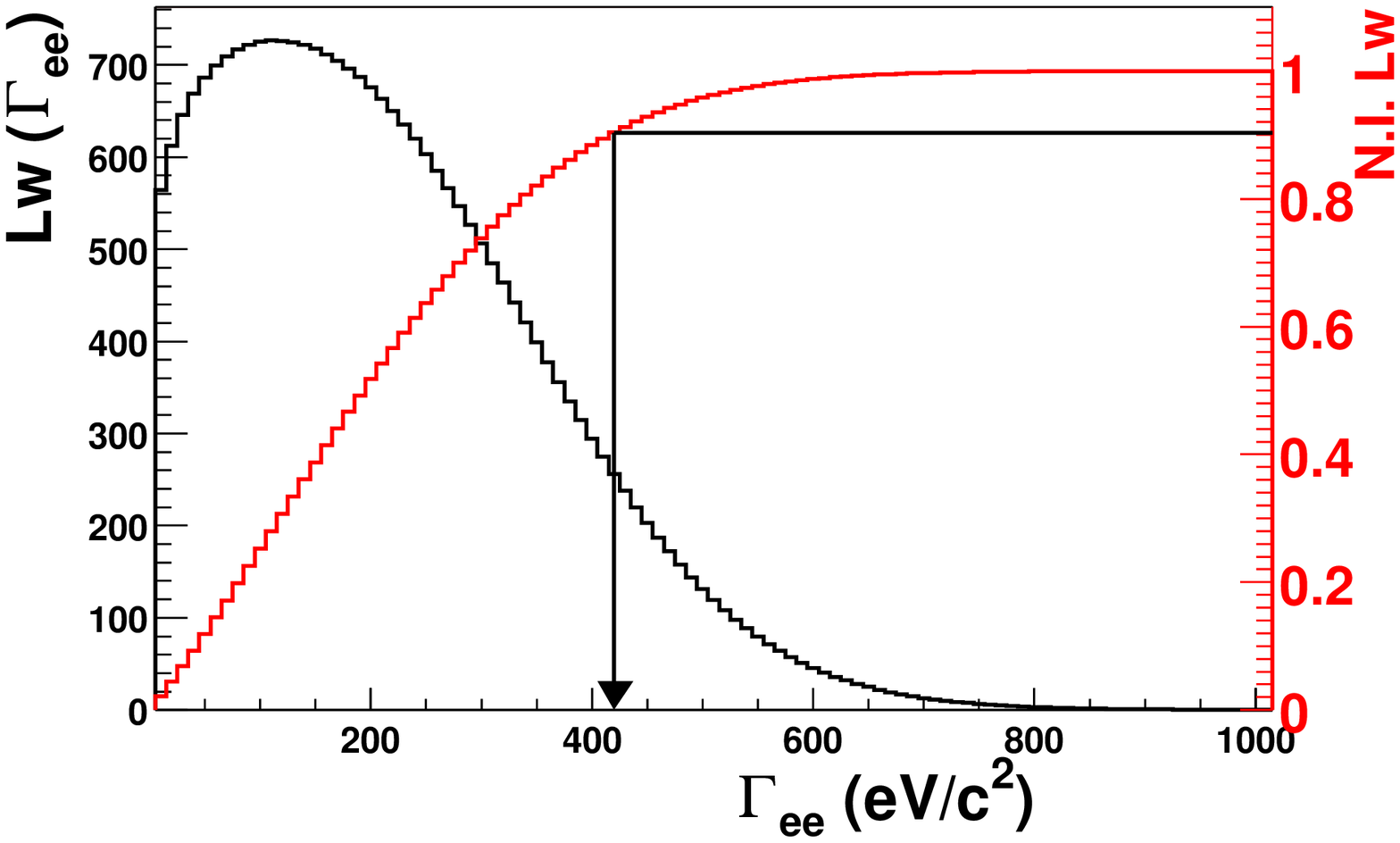}
\centerline{(b) ${\cal L}_w$ w.r.t. $\Gamma_{ee}$}
\end{minipage}
\caption{\label{fig_xsqscan}The weighted likelihood (${\cal L}_w$)
distribution with respect to (w.r.t.) $\Gamma_{ee}$ and $\Gamma_t$ (a); and
to $\Gamma_{ee}$ with $\Gamma_t$ integrated (b). The ``N.I.'' in
(b) indicates normalized integral value for ${\cal L}_w$. }
\end{figure}

For each pair of $\Gamma^i_{ee}(=[i\times 10]$~eV/$c^2$) and 
$\Gamma^j_t(=[20+j\times 8]$~MeV/$c^2$), fitting the $R$ data with
the $\chi^2$ determined from Eq.~\eref{chsqfm}, we obtain a best
estimated $\chi^2$ as a function of the $\Gamma^i_{ee}$ and $\Gamma^j_t$,
or equivalently a relative likelihood, viz. 
 \beq
 {\cal L}_r (\Gamma^i_{ee},\Gamma^j_t)=
 \exp \left(-\frac{1}{2} \chi^2(\Gamma^i_{ee},\Gamma^j_t) \right)~.
 \eeq
Instead of using this ${\cal L}_r$ directly, we further construct a 
weighted likelihood as follows
 \beq
\begin{array}{ll}
 {\cal L}_w (\Gamma^i_{ee},\Gamma^j_t)&=
 f_N \cdot {\cal L}_r (\Gamma^i_{ee},\Gamma^j_t) \times \\
 &{\displaystyle \frac{1}{\sqrt{2\pi} \sigma_{\Gamma_t} }
 \exp \left(-\frac{(\Gamma^j_t-\Gamma_t)^2}{\sigma^2_{\Gamma_t}}
 \right)~},
 \label{wtlklhd}
\end{array} 
 \eeq
where $f_N$ is an arbitrary normalization factor and the Gaussian term 
indicates that the possible total width ($\Gamma^j_t$) of the $\y$
is considered to distribute as a Gaussian with the mean value
$\Gamma_t=88$~MeV/$c^2$ and the standard deviation
$\sigma_{\Gamma_t}=23.8$~MeV/$c^2$, which have been determined by the
$\bbcol$~\cite{babay4260}. The weighted likelihood (${\cal L}_w$) as a 
function of the $\Gamma^i_{ee}$ and $\Gamma^j_t$ of the $\y$ is shown in 
Fig.~\ref{fig_xsqscan}(a). Summing ${\cal L}_w(\Gamma^i_{ee},\Gamma^j_t)$ 
with respect to $\Gamma^j_t$, we
obtain the variation of ${\cal L}_w$ versus $\Gamma_{ee}$ as shown
in Fig.~\ref{fig_xsqscan}(b). The integral of the likelihood curve
gives the upper limit of $\Gamma_{ee}$ of the $\y$ at 90\% confidence 
level (C.L.):
\beq
\Gamma^{\y}_{ee}< 420\mbox{ eV/}c^2~.
\eeq

\begin{figure}[htb]
\includegraphics[width=7.cm,height=6cm]{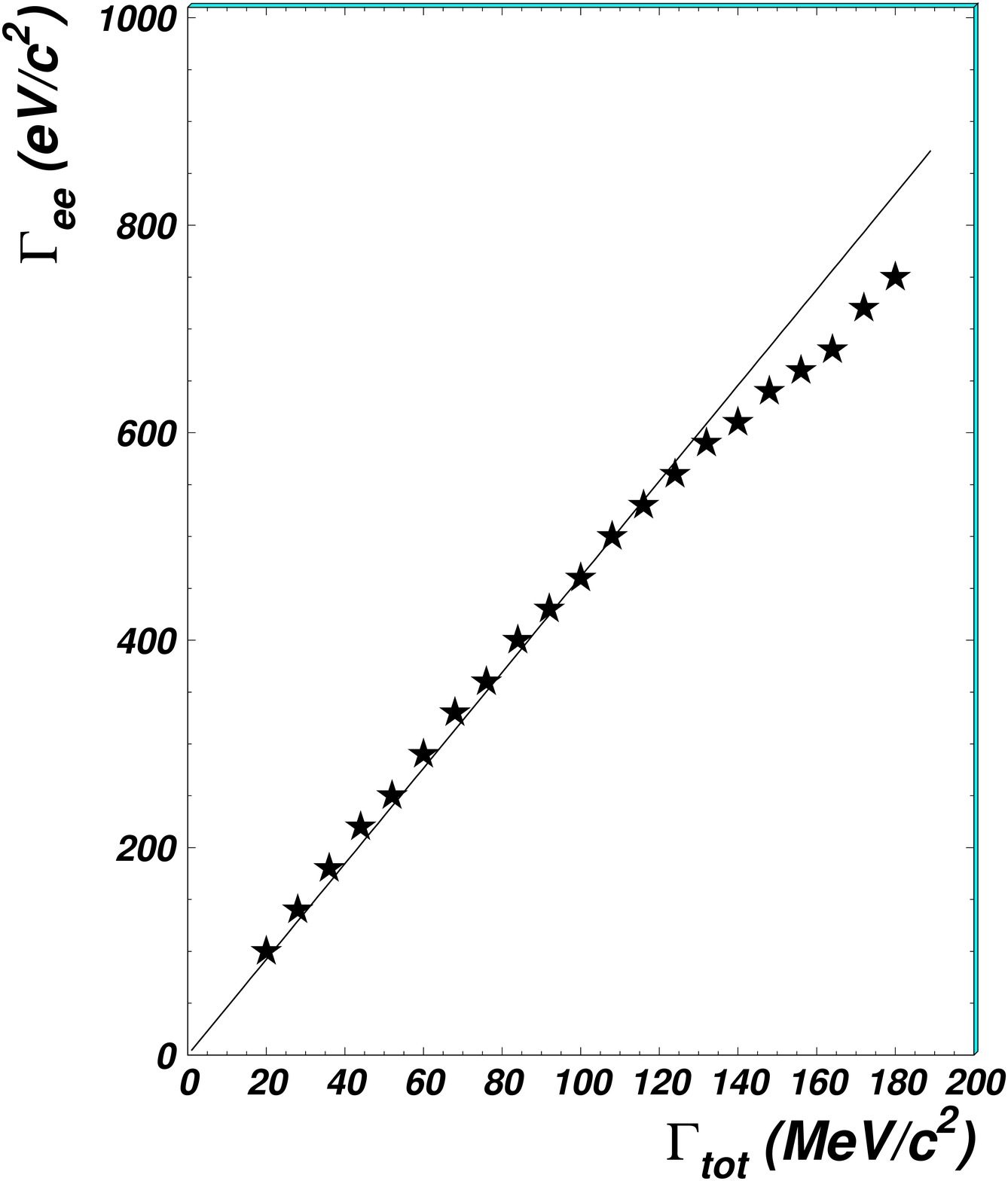}
\caption{\label{fig_gegt} The variation of the upper limit of
$\Gamma_{ee}$ at 90\% C.L. with respect to $\Gamma_t$. The slope
of the solid line denotes the averaged ratio
$\Gamma_{ee}/\Gamma_t=4.6 \times 10^{-6}$.}
\end{figure}

By virtue of Fig.~\ref{fig_xsqscan}(a), we sum up ${\cal L}_w
(\Gamma^i_{ee},\Gamma^j_t)$ with respect to $\Gamma^i_{ee}$ to
90\% fraction of the total area then obtain the upper limit of
$\Gamma^{\y}_{ee}$ at 90\% C.L. for each $\Gamma^{\y}_t$. So we
obtain the variation of the upper limit of $\Gamma_{ee}$ versus
$\Gamma_t$ as shown in Fig.~\ref{fig_gegt}. All the points are
almost in a straight line, which indicates the ratio of the two
quantities, or the upper limit of the branching fraction of $\y\to
\EE$ does not depend on the total width. Taking the slope of the
solid line in Fig.~\ref{fig_gegt}, we get
$$ \BR(\y\to \EE)< 4.6 \times 10^{-6},$$
at 90\% C.L.

\section{Other possibilities}\label{sect_four}

Although it is reasonable to treat all the $\psi$-resonances coherently
as we once mentioned in Sect.~\ref{fitpsisc}, one may argue that other
possibilities could exist since there is limited information about 
the properties of these $\psi$-resonances. 
Thereby it's better to take a variety of effects into consideration.

First, there are three schemes for summation of $\psi$ resonances in 
this analysis: through cross sections, amplitudes, or amplitudes together 
with relative phases, as discussed in Sect.~\ref{fitpsisc}, the 
corresponding results are $\Gamma_{ee}<30$, $10$, $420$ eV$/c^2$ at 90\%
C.L. respectively, where the last one has been given in the previous 
section.

Second, we consider the effect of different background shapes in
the fit. In Sect.~\ref{fitpsisc}, we adopt the first order
polynomial to depict the background, nevertheless, higher order
polynomial can be used to delineate the background as well.

In addition, we notice one background shape once adopted by DASP
group, who tried to take into account the threshold effect of the
charmed mesons~\cite{daspbg},
 \beq
 \sigma_{dasp} =\sigma_{\EE\to\MM} \cdot (A_0 + \sum \limits_{j=1}^{6}
 A_j \beta^3_j F^2 ) ,
 \label{daspfm}
 \eeq
with
 \beq
 F = \frac{1}{1 - s/(3.1 \mbox{ GeV } )^2 }~,
 \label{ffafm}
 \eeq
where $A_0$ describes the contribution from continuum; $j$ ranging from 1
to 6 indicates the $\ddb$, $D \bar{D}^*$, $D^* \bar{D}^*$, $D_s
\bar{D}_s$, $D_s \bar{D}_s^*$, $D_s^* \bar{D}_s^*$ thresholds,
respectively. $A_j$ are free parameters\footnote{
It should be noticed that since the $D_s^* \bar{D}_s^*$ threshold 
has been taken into account in description of the $\y$, the parameter
$A_6$ is set to be zero in the corresponding fit.}; 
$\beta_j$ are the velocities of the relevant particles; 
and $F$ is a oversimplified form factor defined above.

The fit results show that the effect of polynomial background is
at the same level with or smaller than that of the DASP background, so as
an estimation, we adopt the linear and the DASP backgrounds as two
typical cases for background description. Comparing with those
of linear background fit, the results of DASP background fit for  
three summation schemes of $\psi$ resonances are 
$\Gamma_{ee}<50$, $30$, $250$ eV$/c^2$ at 90\% C.L. respectively,

Last, we also consider the possible effect of the fraction $f$ on the
measured $\Gamma_{ee}$ of the $Y(4260)$. Our fits indicate that with 
the increasing of factor $f$ the upper limit of $\Gamma_{ee}$ decreases 
and $vice~versa$. The upper limit for $\Gamma_{ee}$ varies from $580$ 
to $280$ eV$/c^2$ when $f$ changes from $0.3$ to $0.7$.

Taking all the aforementioned possibilities into account, we adopt the
most conservative result as our final estimation, that is
\beq
\Gamma^{\y}_{ee}< 580 \mbox{ eV/}c^2~,
\eeq
at 90\% C.L.
\section{Discussion}

According to our study, the conservative estimation for the upper limit 
of $\Gamma_{ee}$ of the $Y(4260)$ is about 580~eV/$c^2$ at 90\% C.L., which 
is almost two times larger than the estimation presented in 
Eq.~\eref{sygee}.

Utilizing our upper limit $\Gamma_{ee}<580$~eV/$c^2$ for the
$Y(4260)$, together with the relation of Eq.~\eref{gambr}, we
obtain the lower limit of the branching fraction at 90\% C.L. to be
$$ \BR(Y(4260) \to \ppjpsi) > 0.58\%~.$$
This means that the partial width $\Gamma(Y(4260) \to
\ppjpsi) \ge 508$~keV/$c^2$ at 90\% C.L., which is much larger
than the corresponding partial widths of $\psp$
(89.1~keV/$c^2$)~\cite{pdg04} and $\pspp$
(44.6~keV/$c^2$)~\cite{pdg04,cleocppjp}.

CLEOc measured the cross section for $\ppjpsi$ channel to be
58 pb$^{-1}$~\cite{cleoy4260}, which is consistent with the BaBar result, 
50 pb$^{-1}$~\cite{babay4260}. As an estimation, 
we regard the central value calculated in Eq.~\eref{gambr} to be the same 
for CLEOc and BaBar, but adopt the improved accuracy provided by CLEOc, 
then we can obtain the following lower limits at 90\% C.L.
\begin{eqnarray*}
\BR(Y(4260) \to \ppjpsi)        &>&0.6\%~, \\
\BR(Y(4260) \to \piz \piz \jpsi)&>&0.2\%~, \\
\BR(Y(4260) \to K^+ K^- \jpsi)  &>&0.1\%~, \\
\end{eqnarray*}
and
$$ \BR(Y(4260) \to X \jpsi) > 1.3 \%~, $$
where $X$ denotes $\pp$(37), $\piz\piz$(8), $\kk$(3), 
$\eta$(5), $\piz$(1), $\etap$(0), $\pp\piz$(0), and $\eta\eta$(1).
Here the numbers of the observed events by CLEOc are
presented in parentheses.

As we notice up to now only results on hidden-charm final states were
reported about the $\y$; measurements involving open-charm are anxiously 
awaited to confirm existing speculations or provide clues for further 
theoretical inquiry. In addition, more accurate $\Gamma_{ee}$ is still
needed for a better understanding of the nature of the $\y$.
 
\section*{Acknowledgments}
We would like to thank Dr. H.B. Li for friendly discussion.

\end{document}